\documentclass[pdfusetitle,aps,reprint,prl,twocolumn,superscriptaddress,preprintnumbers]{revtex4-1}
\pdfoutput=1
\usepackage{amssymb}
\usepackage{amsmath}
\usepackage{epsfig}
\usepackage{hyperref}
\usepackage{breakurl}
\usepackage{multirow}
\usepackage{array}
\usepackage{bm}
\usepackage{color}
\usepackage{tikz}
\usepackage[utf8]{inputenc}
\usepackage{braket}
\usepackage{graphicx}

\usetikzlibrary{decorations.markings}
\usepackage{cleveref}
\crefformat{section}{\S#2#1#3} 
\crefformat{subsection}{\S#2#1#3}
\crefformat{subsubsection}{\S#2#1#3}

\makeatletter
\def\simgt{\mathrel{\lower2.5pt\vbox{\lineskip=0pt\baselineskip=0pt
           \hbox{$>$}\hbox{$\sim$}}}}
\def\simlt{\mathrel{\lower2.5pt\vbox{\lineskip=0pt\baselineskip=0pt
           \hbox{$<$}\hbox{$\sim$}}}}

\def\sectionskip{\vskip .2 cm}

\makeatother

\def\spa#1.#2{\left\langle#1\,#2\right\rangle}
\def\spb#1.#2{\left[#1\,#2\right]}
\def\sand#1.#2.#3{%
\left\langle#1{\vphantom1}\right|{#2}\left|#3\right]}%
\def\sandmp#1.#2.#3{%
\left\langle#1{\vphantom1}\right|{#2}\left|#3\right]}%
\def\sandpm#1.#2.#3{%
\left[#1{\vphantom1}\right|{#2}\left|#3\right\rangle}%
\def\sandmm#1.#2.#3{%
\left\langle#1{\vphantom1}\right|{#2}\left|#3\right\rangle}%
\def\sandpp#1.#2.#3{%
\left[#1{\vphantom1}\right|{#2}\left|#3\right]}%

\renewcommand{\imath}{\mathrm{i}}

\def\Section#1{\noindent {\it #1}}
\newcommand{\be}{\begin{equation}}
\newcommand{\ee}{\end{equation}}

\def\S{{\mathbb S}}

\begin{document}

\title{  Dynamics for  Super-Extremal Kerr Binary Systems   at   ${\cal O}(G^2)$}

\author{Yilber Fabian Bautista}
\affiliation{Institut de Physique Théorique, CEA, Université Paris–Saclay,
F–91191 Gif-sur-Yvette cedex, France}

\begin{abstract}
Using the recently derived higher spin gravitational Compton amplitude from  low-energy  analytically continued ($a/Gm\gg1$) solutions of the Teukolsky equation for the scattering of a gravitational wave off the  Kerr black hole,   observables for non-radiating
super-extremal Kerr binary systems at second   post-Minkowskian (PM) order and up to sixth order in spin are computed.  The relevant 2PM amplitude is obtained from the triangle-leading singularity  in conjunction with  a generalization of the holomorphic classical limit   for massive particles with spin oriented in generic directions. 
Explicit results for the 2PM eikonal phase written for  both Covariant and Canonical spin supplementary conditions -- CovSSC and CanSSC respectively -- as well as the 2PM    linear impulses and individual spin kicks in  the CanSSC are presented. The observables  reported in this letter are expressed in terms of   generic contact deformations of the gravitational Compton amplitude, which can then be specialized to Teukolsky solutions. In the latter case, the resulting   2PM  observables break the newly proposed spin-shift symmetry of the 2PM amplitude  starting   at the fifth order in  spin.  Aligned spin checks as well as the  high energy behavior of the computed  observables are discussed.

\end{abstract}

\maketitle

\sectionskip
\Section{Introduction.}
The application of  QFT-inspired methods to  compute    observables in  classical   gravity has seen  tremendous  advances in the last years \cite{Cheung:2018wkq,Kosower:2018adc,Bern:2022jvn,Bern:2021yeh,Bern:2021dqo,Bern:2020buy,Bern:2019nnu,Dlapa:2022lmu,Dlapa:2021npj,Kalin:2020fhe,Mogull:2020sak,Jakobsen:2022psy,Elkhidir:2023dco,Brandhuber:2023hhy,Herderschee:2023fxh,Georgoudis:2023lgf,DiVecchia:2021bdo} due to their potential   relevance for   analyzing  measured signals  in gravitational wave (GW)  detectors \cite{LIGOScientific:2016aoc}. The use of these techniques is justified by   the separation of scales that allows  to treat  physics  problems in an effective manner. For instance, an isolated black hole (BH) seen from  far away can be thought of as a point particle, and its finite size effects such as spin multipole moments can be modeled by effective operators in (classical) EFT constructions \cite{Levi:2015msa,Bern:2022kto,Chung:2020rrz,Jakobsen:2022fcj,Liu:2021zxr,Arkani-Hamed:2017jhn,Chung:2018kqs,Bautista:2022wjf,Aoude:2022trd,Aoude:2022thd}. These
operators are accompanied by free coefficients  parametrizing  the UV ignorance of the effective model;   they can be   fixed by matching computations in the effective and the full theory.  Following this logic, an astonishingly  simple description of an isolated linearized  Kerr  BH \cite{Vines:2017hyw} -- effectively a super-extremal (SE) Kerr BH with spin parameter  $a^\star=\frac{a}{Gm}\gg1$, where $m$ and $a$ are its mass and ring radius respectively --  as an elementary particle of infinite spin minimally  coupled  to gravity, has recently appeared  in the literature \cite{Guevara:2018wpp,Chung:2018kqs,Arkani-Hamed:2019ymq,Aoude:2020onz}.  

Actual Kerr BHs are  however  neither isolated objects in nature nor linearized solutions of the field equations as they possess   spin parameters satisfying  $a^\star \le1$. A correct description of these objects requires  therefore
resummations of the perturbative computations.
Accomplishing these resummations  from a perturbative amplitude approach is a heroic task    not yet  addressed   along these  lines of reasoning, and therefore relying on   the effective one-body  formalism \cite{Buonanno:1998gg,Buonanno:2000ef,Damour:2001tu,Vines:2017hyw,Buonanno:2005xu,Khalil:2023kep},  as an  alternative way   to study these more realistic Kerr BHs scenarios
\footnote{In principle, in an EFT approach, a spin resummation  prescription can  be obtained by allowing the  effective coefficients to be functions of $a^\star$ \cite{Bautista:2022wjf,Saketh:2022wap}.  }. 
Nevertheless, perturbative approaches are still very useful to gain insights into  the physics of actual  Kerr BH systems, and in this work, we follow these lines to  study systems made  not of actual Kerr BHs, but of their close relatives, SE Kerr BHs (fastly rotating  objects).

Although   SE Kerr BHs are not real objects in nature, 
they share many features of actual Kerr BHs, as the two are related via  analytic continuation of the spin parameter from the physical region $a^\star \le1$, to the SE $a^\star\gg1$  region. For observables that cannot probe the nature of the BH horizon -- which we shall refer to as  \textit{ true conservative} observables \footnote{For the systems considered in this work, we discard  radiative effects encoded  in the emission of gravitational waves towards future null infinity. } -- this continuation should not possess any subtlety and their values 
 computed for one kind of objects or the others should coincide in the overlapping (continuation) region \footnote{At low spin orders, this continuation is equivalent to the spin multipole expansion at fixed order in $G$.   }. In this sense, SE Kerr  observables  readily encapsulate part of the  dynamic for actual Kerr BHs \cite{Vines:2017hyw,Bini:2017xzy,Chung:2020rrz,Bern:2020buy,Kosmopoulos:2021zoq,Chen:2021kxt,Jakobsen:2022fcj,Levi:2016ofk,Levi:2015msa,Levi:2022rrq,Kim:2022pou,Liu:2021zxr,Jakobsen:2022fcj,Goldberger:2017ogt,Jakobsen:2021lvp,Jakobsen:2021zvh,Jakobsen:2022psy,Jakobsen:2022zsx,FebresCordero:2022jts,Blanchet:2013haa,Porto:2016pyg,Levi:2018nxp,Levi:2015ixa,Levi:2016ofk, Levi:2020kvb, Antonelli:2020aeb,Levi:2020uwu, Antonelli:2020ybz, Kim:2021rfj,Levi:2019kgk,Levi:2020lfn,Maia:2017gxn,Cho:2022syn,Cho:2021mqw,Kim:2022pou,Kim:2022bwv,Levi:2022dqm,Levi:2022rrq}. Observables that can probe the nature of the  BH horizon -- \textit{absorptive} observables accounting for  fluxes of energy 
 at the BH horizon --
  are more subtle since   the notion of absorption does not exist for objects without a horizon; hence, SE Kerr BH observables look always \textit{ conservative}. These can however receive  contributions from effective operators that mimic the physical effects happening at the  horizon of an actual Kerr BH, but whose definite identification requires comparison to alternative approaches to study absorptive effects \cite{Tagoshi:1997jy,Poisson:2004cw,Chatziioannou:2012gq, Chatziioannou:2016kem,Isoyama:2017tbp,Goldberger:2005cd,Goldberger:2020fot,Porto:2007qi, Saketh:2022xjb,Saketh:2022xjb}.

\begin{figure}
    \centering
    \includegraphics[scale=0.23]{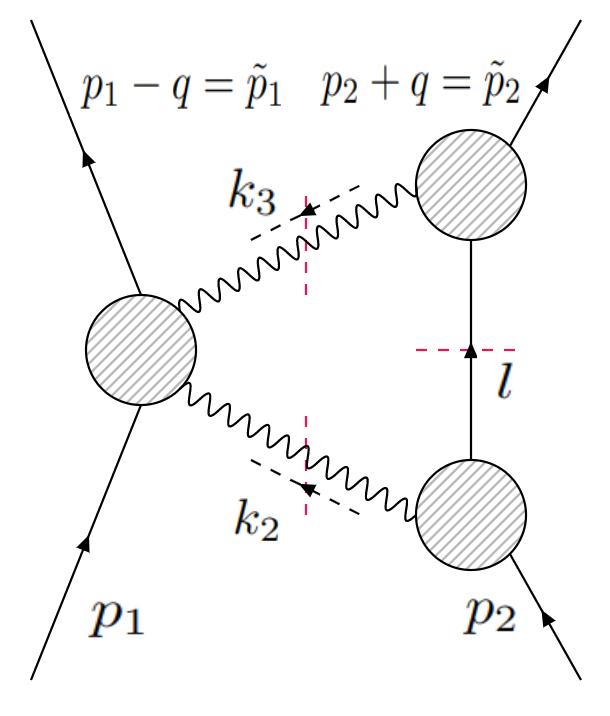}
    \vspace{-0.7cm}
    \caption{Triangle leading-singularity configuration \cite{Cachazo:2017jef}.}
    \label{leading-singularity}
\end{figure}

In this work, we  compute \textit{conservative} observables for the  scattering of two SE Kerr BHs at $\mathcal{O}(G^2)$, but  whose content can potentially encode \textit{true conservative} as well as \textit{absorptive} effects for actual Kerr BHs. For this, we use the recently extracted higher spin gravitational Compton amplitude from low-energy solutions of the Teukolsky equation in the SE   region \cite{Bautista:2022wjf}.  The observables of interest in this work are the linear impulse and the individual spin kicks, for generic spin orientations. We will use the eikonal phase as an intermediate object, and compute the observables using formula \eqref{eq:master_observable}, proposed by the authors of references \cite{Bern:2020buy,Kosmopoulos:2021zoq}. Since contact deformation of the Compton amplitude  enters the 2PM amplitude only through the triangle diagrams (\cref{leading-singularity}), we expect this formula \eqref{eq:master_observable} (showed to be valid for orders $a^{n\le4}$ at 2PM \cite{Bern:2020buy,Kosmopoulos:2021zoq,Chen:2021kxt}) to continue to be valid for higher spin orders.

\sectionskip
\Section{A tree-level gravitational Compton amplitude from the Teukolsky equation.}
In \cite{Bautista:2022wjf}, an ansatz for the $\mathcal{O}(G)$  opposite helicity,  higher spin gravitational Compton amplitude  of the form (momentum cons.  $p_4 = p_1 + k_2 + k_3$)
\begin{equation}\label{eq:ansatzspin}
  A_4^{(S)}  =  A_4^0 \left(e^{(2w+k_3-k_2)\cdot a} {+} P_\xi (k_2\cdot a,{-}k_3 \cdot a, w\cdot a) \right )_{2S}\,,
\end{equation}
was written invoking physical constraints such as   locality, unitarity, 3-point factorization,and crossing symmetry, together with a prescription to write  contact deformations -- capture by $P_\xi$ --  that match Teukolsky solutions only in a non-trivial manner \footnote{A more general ansatz for a gravitational Compton amplitude including  neutron stars was provided recently in \cite{Haddad:2023ylx}.  Other  approaches to computing the higher spin gravitational Compton amplitude have been explored in the literature \cite{Aoude:2022trd,Chiodaroli:2021eug,Bern:2022kto,Aoude:2022trd,Aoude:2022thd,Haddad:2023ylx,Cangemi:2022bew,Bjerrum-Bohr:2023jau,Alessio:2023kgf}  }. 

The scalar contribution, $ A_4^0$ in \eqref{eq:ansatzspin}, encodes the helicity  and physical pole structure of the  amplitude, whereas the terms inside the big parenthesis  are  only functions of  the kinematic invariants and the spin of the massive legs. Explicitly, the former reads 
\vspace{-0.2cm}
\begin{equation}
\label{eq:compt_scalar_hcl}
     A_4^{0}= 32\pi G m^2\frac{(\epsilon_2\cdot u)^2(\tilde{\epsilon}_3\cdot u)^2 }{ \xi },\,\,\, \xi  = \frac{(s-m^2)^2}{m^2t}\,,
\end{equation}
which we  choose to evaluate  in the gauge 
\begin{equation}\label{eq:wdef}
    \epsilon_2=\frac{\sqrt{2} |3]\langle 2|}{[32]} \propto \tilde{\epsilon}_3 = \frac{\sqrt{2} |3]\langle 2|}{\langle32\rangle }\,,\,\,\,w^\mu := \frac{u\cdot k_2}{u\cdot \epsilon_2} \epsilon^\mu_2\,.
\end{equation}
Here we have   included the    massless vector $w^\mu$  entering in \eqref{eq:ansatzspin}, which is   constructed from the spinors of the massless legs, and $u=p_1/m$, is the incoming massive leg's four-velocity, with $m$ its respective mass.  

The   function $ P_\xi$ contains   contact deformations of the BCFW exponential -- some of which at  the same time cure the unphysical singularities  that appear starting at  $ \mathcal{O}( a^5)$ when the exponential function is expanded -- was   written as a Laurent expansion in the optical parameter $\xi$, having the  explicit form    given  in \eqref{eq:ansatz}.  This expansion was further   parametrized by three    multi-variable polynomials  $p^{(m)}_{|a|},q^{(m)}_{|a|},r^{(m)}_{|a|}$, functions of  $(k_2\cdot a, -k_3 \cdot a, w\cdot a) $, symmetric in their first two arguments, but  also including a linear correction in $\omega |a|$, with $\omega\approx (s-m^2)/2m$,  the energy of the massless legs.
Up to the sixth order in spin, these polynomials have the explicit form  given by eqs. (\ref{eq:polr}-\ref{eq:polpq}),  and  contain two type of spin operators:
 \textit{regular} operators, functions of only $\mathcal{R}= \{ k_2\cdot a,k_3\cdot a,w\cdot a \}$, and \textit{ exotic} operators, functions of   $\mathcal{R} \cup\{\omega|a|\}$.  The former will encapsulate the real contributions 
 (at the level of the phase shift) to the solution to the Teukolsky equation, whereas the latter accounts for contributions that are  imaginary  when the BH rotation parameter satisfies the inequality $a^\star\le1$.

Each effective operator in  $p^{(m)}_{|a|},q^{(m)}_{|a|},r^{(m)}_{|a|}$ is accompanied by  a free coefficient $c_{i}^{(j)},d_i^{(j)},f_i^{(j)}$ respectively;  contact deformations were shown to appear  starting  at  fourth order in spin as known from the work  \cite{Chung:2018kqs}.   These free coefficients  were further fixed by requiring that \eqref{eq:ansatzspin} matches the $\mathcal{O}(G)$ sector  of the low energy limit  ($\epsilon=Gm\omega\ll1$) of the  scattering  amplitude   for the scattering   of a GW off the Kerr BH,  computed with  the tools of black hole perturbation theory   \cite{Sasaki:2003xr,Bautista:2021wfy}. Explicit solutions up to  six order in spin can be found  in Table 1 in \cite{Bautista:2022wjf}, which we include in \cref{tab:Teukolskysolutions} of Appendix B,  for the reader's convenience. Remarkably, up to the fourth order in spin, the Teukolsky solution perfectly matches the classical limit of the  undeformed minimal coupling gravitational Compton amplitude of Arkani-Hamed,  Huang, and  Huang \cite{Arkani-Hamed:2017jhn}, given by the expansion of the exponential in \eqref{eq:ansatzspin}, up to  $\mathcal{O}( a^4)$.  Up to this order, Teukolsky solutions are polynomials in $a^\star$, therefore, providing a unique answer for the analytically continued Kerr results to the SE  Kerr approximation $a^\star \gg1$.   Starting at the fifth order in spin, Teukolsky solutions contain complex,  non-rational functions of $a^\star$, which are in addition discontinuous at $a^\star=1$. Therefore, a prescription for analytic continuation to the SE region was needed. The two different  prescriptions provided in \cite{Bautista:2022wjf}  were labeled by the parameter $\eta$, which takes values $\pm1$, with the sign determined by the continuation procedure. Non-trivially, contributions in the Teukolsky solutions that were real before the continuation   uniquely fix the free coefficients for \textit{regular} operators in \eqref{eq:ansatzspin}, and are independent of such continuation prescriptions, whereas the pieces that were imaginary before the continuation,  fix the coefficients of \textit{exotic} operators,  modulo a sign. 
The \textit{exotic} contributions are believed to  encode only physical effects happening   at the BH horizon for actual Kerr  ($a^\star \le1$)  systems, but their precise physical  interpretation in a  realistic Kerr context  is beyond the scope of this work. We refer the reader to  the recent work \cite{Saketh:2022xjb} for a related analysis of  horizon dissipation for Kerr systems (see also  \cite{Tagoshi:1997jy,Poisson:2004cw,Chatziioannou:2012gq, Chatziioannou:2016kem,Isoyama:2017tbp,Goldberger:2005cd,Goldberger:2020fot,Porto:2007qi, Saketh:2022xjb}).  

Some comments from this matching are in order:
1) After analytic continuation, imaginary contributions become real, therefore providing \textit{conservative} information for the SE Kerr  system. This is  a consequence of erasing the BH horizon in the continuation procedure, therefore, removing any source of \textit{ dissipation}. 
2) In the matching of \eqref{eq:ansatzspin} to  the low energy limit of Teukolsky solutions, terms of the form $\epsilon^n (a^{\star })^m f(a^\star) $ for $m\ne n $ and $f$ and  function of the rotation parameter,  were discarded as they do not contribute to the tree-level amplitude. Similarly, terms of the form $\epsilon^n \log \epsilon\, (a^{\star })^m f(a^\star) $ that do not produce $\mathcal{O}(G)$ contributions were removed.
These terms might however  become important for a Compton amplitude that matches the actual Kerr ($a^\star \le1$) solutions \footnote{ Table 2 in \cite{Bautista:2022wjf} contain coefficients that match the Compton ansatz \eqref{eq:ansatzspin} to  low energy  Kerr solutions without analytic continuations, but have  however discarded  $\epsilon^n \log \epsilon \,(a^{\star })^m f(a^\star) $  terms. } . 

For the same helicity sector, the $\mathcal{O}(G)$  Teukolsky solutions were shown to match spectacularly the analogous exponential $ \tilde{A}_0 e^{-(k_2+k_3)\cdot a}$, with $\tilde{A}_0$ the spin independent contribution, with checks made  up to $\mathcal{O}(a^6)$ in the SE  region. The results were also shown to be  independent of the  continuation prescription.

\sectionskip
\Section{Leading PM   eikonal phase for super-extremal  Kerr.} In this work we are interested in computing  canonical observables for binary systems  at  the second PM order.  Binary   2PM    observables however will necessarily require 1PM information entering   as iteration terms in the operator formulation \cite{Kosower:2018adc}, 
the 2PM Hamiltonian \cite{Bern:2020buy,Chen:2021kxt}, or equivalent as quadratic  eikonal  contributions in the  formula \eqref{eq:master_observable} for the computation of canonical observables directly  from the eikonal phase \cite{Bern:2020buy,Kosmopoulos:2021zoq}. Driven by this, in this section we revisit the computation of the 1PM eikonal phase for the scattering of two SE Kerr BHs to all orders in  spins. We start by recalling the tree level, all orders in spin two-body amplitude is obtained from  the unitarity gluing of two SE Kerr 3-point amplitudes \cite{Guevara:2018wpp,Chung:2018kqs,Vines:2017hyw}, resulting in the compact expression \cite{Chung:2020rrz}:
\begin{equation}\label{eq:1PMbare}
M^{(\text{1PM})}_{\text{bare}}{=} {-}\frac{16\pi G\,(m_1 m_2)^2}{q^2}\cosh\left( 2\theta{+}i\frac{\mathcal{E}_1/m_1{+}\mathcal{E}_2/m_2}{m_1m_2\sinh(\theta)}\right)\,.
\end{equation}
Here we have used  the notation $\mathcal{E}_i=\epsilon_{\mu\nu\rho\sigma}q^\mu p_{1}^\nu p_{2}^\rho s_{i}^\sigma$, where $p_i,s_i$ are the respective momenta and spins of the incoming BHs, and $q$ is the momentum transfer. The     Lorentz boost factor was introduced via the hyperbolic functions
$
\cosh(\theta)=\sigma=\frac{p_1{\cdot}p_2}{m_1m_2}\,,\,\,\,\sinh(\theta)=\sigma v\,,
$ 
with $v$ the two bodies' relative velocity. The ``bare" label in \eqref{eq:1PMbare} indicates massive  spin polarization tensors have been removed, and the  observables computed with this prescription have 
the rotational gauge freedom fixed by the  Tulczyjew-Dixon  covariant spin supplementary condition  (CovSSC) $p_{b}S^{ab}=0$ \cite{Tulczyjew,Tulczyjew2}.   Hamiltonian observables      however are customarily computed in the canonical  Newton-Wigner
SSC (CanSSC)  $S^{ab}(p_b+\sqrt{p^2}\delta_{0b})=0$, with $\{a,b\}$ local frame indices \cite{RevModPhys.21.400,Levi:2015msa}\footnote{With an abused of notation, we have used the same symbol for denoting the spin tensor entering in observables computed using   either the CovSSC or the CanSSC. Each object should be easily distinguished by setting the  
$\tau$ parameter to $0$ or $1$ respectively.}. A way to  make the previous amplitude  satisfy the latter constraint is provided by 
  dressing the bare amplitude with the Thomas-Wigner rotation factors  \footnote{More precisely, this corresponds to a canonical  alignment of the incoming and outgoing massive polarization tensors through a Lorentz transformation \cite{Chung:2019duq,Chung:2020rrz}. },
\begin{equation}\label{eq:1PMdressed}
M_{\text{dressed}}^{\text{(nPM)}}= M^{(\text{nPM})}_{\text{bare}} U_1 U_2\,,\,\,\, U_i = e^{\frac{i\tau\mathcal{E}_{i}}{Em_i(E_i+m_i)}}\,,\,\,\,n =1,2.
\end{equation}
where $E{=}E_1{+}E_2$, is the sum of the individual  bodies' energies, and $\tau{=}1$, is a parameter that keeps track of the CanSSC prescription. These rotation factors are written in the center of mass (CoM) frame, where the momenta of the BHs are parametrized in the following way  (see $e.g.$  \cite{Bern:2020buy} for details):
\begin{equation}
p_1{=}{-}(E_1,\boldsymbol{p})\,,\,\,p_2{=}{-}(E_2,-\boldsymbol{p})\,,\,\,q{=}(0,\boldsymbol{q})\,,\,\,\boldsymbol{p}{\cdot}\boldsymbol{q}{=}\frac{\boldsymbol{q}^2}{2}\,,
\end{equation}
Here, $\boldsymbol{p}$ is the asymptotic  incoming three-momentum -- sometimes also referred to as $\boldsymbol{p}_\infty$ -- and $\boldsymbol{q}$ is the three-momentum transfer in the scattering process.  The covariant spin operators can analogously be mapped to  their CoM representations via  
\be\label{eq:spin_kin_com}
\begin{split}
\mathcal{E}_i &= E\,\boldsymbol{s}_i\cdot(\boldsymbol{p}\times\boldsymbol{q})\,,\,\,\,
q\cdot s_i=\boldsymbol{q}\cdot\boldsymbol{s}_i+\mathcal{O}(\boldsymbol{q}^2)\,,
\\
|s_i|&=|\boldsymbol{s}_i|\,,\,\,p_i\cdot s_j=\epsilon_{ij} \frac{E}{m_i} \boldsymbol{p}\cdot \boldsymbol{s}_j\,,\,\, \epsilon_{12}=-\epsilon_{21}=1\,.
\end{split}
\ee
The eikonal phase for this $2\to2$ scattering  process is then obtained from the 2-dimensional Fourier  transform of the two-body amplitude into impact parameter space. For the generic 1PM and 2PM cases, we have \cite{AMATI198781,Melville:2013qca,Akhoury:2013yua,DiVecchia:2019myk}
\begin{equation}\label{eq:eikonalgen}
\chi^{\text{(nPM)}}_\tau{= }\frac{1}{4m_1 m_2 \sqrt{\sigma^2-1}}\int\frac{d^2\boldsymbol{q}}{(2\pi)^2} e^{-i\boldsymbol{q}{\cdot} \boldsymbol{b}} M_{\tau}^{\text{(nPM)}}\,,\,n =1,2,
\end{equation}
with  $M_{\tau}^{\text{(2PM)}}$ computed from  the triangle leading singularity (LS)  \cref{leading-singularity}. Here we have left explicit the $\tau$ label which for $\tau=0$, we input the bare (CovSSC) amplitude into \eqref{eq:eikonalgen}, whereas for $\tau=1$,  the Thomas-Wigner rotation factors (CanSSC) need to be supplemented.   

Going back to the 1PM analysis, the  spin operators entering in \eqref{eq:1PMbare} and  \eqref{eq:1PMdressed} simply become  shifts of the impact parameter when the explicit evaluation of   \eqref{eq:eikonalgen} is performed. At the 1PM order, one arrives at the all-spin eikonal phase  
\begin{equation}\label{eq:1pmeik}
\chi_\tau^{\text{(1PM)}}=-\frac{m_1m_2G}{2\sqrt{\sigma^2-1}}
\sum_{\pm}(\cosh(2\theta)\pm\sinh(2\theta))\log(\boldsymbol{b}_\tau^{(\pm)2})\,,
\end{equation}
where $\boldsymbol{b}_\tau^{(\pm)}=\boldsymbol{b}+\sum_{i=1,2}\left(\frac{-\tau}{E_i+m_i}\pm\frac{ E}{m_1 m_2 \sinh(\theta)}\right) \frac{\boldsymbol{p}\times \boldsymbol{s}_i}{m_i} $, and the $\pm$ sum remembers the  helicity sum for the exchanged graviton. 
Let us stress this formula for the eikonal is  valid for generic spin orientations and recovers  previous lower spin results \cite{Bern:2020buy,Kosmopoulos:2021zoq,Chung:2020rrz}.

\sectionskip
\Section{2PM amplitude from the Triangle LS and the non-aligned   HCL.}
Having the 1PM eikonal at our disposal, the next ingredient to compute 
2PM observables via \eqref{eq:master_observable}, is the 2PM eikonal phase for compact objects with generic spin orientations. The relevant one-loop amplitude entering in \eqref{eq:eikonalgen} is controlled by the LS (see \cref{leading-singularity}):
\begin{equation}\label{LS} 
\frac{1}{8m_2\sqrt{-t}} \int_{\Gamma_{\rm{LS}}} \frac{dy}{2\pi y} A_4^{(a_1)}(A) A_3^{(a_2)}(B)  A_3^{(a_2)}(C) ,
\end{equation}
where the  contour $\Gamma_{\text{LS}}$ computes the residue at $y = 0$ minus that at $y= \infty$ \cite{Cachazo:2017jef}.
The momentum labels of the building tree-level blocks are $\{A\}=\{p_1,{-}\tilde{p}_1,k_2^{+},k_3^{-}\}$, $\{B\}=\{p_2,{-}l,{-}k_2^-\}$ and $\{C\}=\{-\tilde{p}_2,l,{-}k_3^+ \}$; namely,  only the opposite helicity configuration of the Compton amplitude will be relevant for  super-extremal Kerr observables 
\footnote{It can be easily shown that for the same helicity sector for the  Compton, the integrand in \eqref{LS} factorizes in the form $A_{4}^{(0)}A_{3}^{(0)}A_{3}^{(0)} e^{-(k_2+k_3)\cdot(a_1+a_2)}\to A_{4}^{(0)}A_{3}^{(0)}A_{3}^{(0)} e^{-q{\cdot}(a_1+a_2)} $, where we have used   \eqref{eq:spin_op} to rewrite the exponents. That is, the spin-dependent part is independent of the loop integration variable $y$, and the LS \eqref{LS} reduces to that for the scalar case which evaluates to zero as shown in the seminal work of Guevara \cite{Guevara:2017csg}. Amplitudes for neutron stars on the other hand  receive contributions from both helicity configurations of the  Compton amplitude \cite{Chen:2021kxt,Aoude:2022thd}.  }.

We use the holomorphic classical limit  (HCL) parametrization for the non-aligned spin scenario discussed in Appendix A,  as an alternative construction of the 2PM amplitude to the usual -- but very related -- unitarity methods \cite{Forde:2007mi,Badger:2008cm,Bern:2020buy,Chen:2021kxt,Aoude:2022trd}. An advantage of the LS construction however is that the classical limit of the triangle graph can be taken from the beginning of the computation. 
For the scalar contributions, the HCL parametrization is the usual one \cite{Guevara:2017csg,Guevara:2018wpp,Bautista:2022wjf}. In the gauge \eqref{eq:wdef},  the scalar Compton amplitude  \eqref{eq:compt_scalar_hcl} has the HCL form 
$
       A_4^{0}(A)=\pi G m_1^2\frac{(2y-v(1+y^2))^4\sigma^2}{v^2y^2(1-y^2)^2}\,,
$
 whereas for the product of the two  3-point amplitudes we have
$
     A_3^{0}(B) A_3^{0}(C) = 8\pi G m_2^4\,.
$
The non-aligned HCL form for the spin-contributions to the 2PM amplitude  becomes however very different from their aligned spin construction. The massless momenta hitting the spin vectors $a_1^\mu$ in \eqref{eq:ansatzspin}, and $a_2^\mu$ in the 3-point amplitudes  $e^{-k_2{\cdot}a_2}\times e^{k_3{\cdot}a_2}$, are now given by
\be\label{eq:spin_op}
\begin{split}
k_2^\mu&{=}\frac{|q|\sqrt{z^2{-}1}(m_2p_1^\mu{-}m_1\sigma p_2^\mu){+}m_1m_2\sigma v q^\mu{+}iz \mathcal{E}^\mu}{2m_1m_2\sigma v}\,,\\
k_3&{=}{-}\frac{|q|\sqrt{z^2{-}1}(m_2p_1^\mu{-}m_1\sigma p_2^\mu){-}m_1m_2\sigma v q^\mu{+}iz \mathcal{E}^\mu}{2m_1m_2\sigma v},\\
w^\mu&{=}{-}\frac{|q|\sqrt{z^2{-}1}\big(m_2p_1^\mu z{+}m_1\sigma p_2^\mu(v-z)\big){+}i(z^2-1)\mathcal{E}^\mu}{2m_1m_2(1{-}vz)\sigma}\,,
\end{split}
\ee
with $z=\frac{1+y^2}{2y}$ and $\mathcal{E}^\mu=\epsilon^{\alpha\beta\gamma\mu} q_\alpha p_{1\beta}p_{2\gamma}$. These expressions contain the leading in $|q|=\sqrt{-q^2}$ contribution to the classical amplitude, therefore discarding unnecessary quantum information before loop integration. 
Notice in particular, the combination $k_2^\mu-k_3^\mu$ is independent of the transfer momentum $q^\mu$; this will be relevant  when we discuss below some caveats of the aligned spin constructions of \cite{Guevara:2018wpp,Bautista:2022wjf} for the computation of the aligned spin scattering angle for the higher  spin $(a^{n>4})$ cases.

Having at hand the  non-aligned HCL parametrization of the building blocks entering in  \eqref{LS}, together with the HCL form of the optical parameter $ \xi \rightarrow - \sigma^2 v^2 \frac{(1-y^2)^2}{4y}$, it is now an easy task to compute the LS \eqref{LS} using the Compton amplitude \eqref{eq:ansatzspin}, since the problem has been reduced to a simple  residue calculation.  We  organize the result of the residue evaluation as follows: Given the set of spin operators $\mathcal{H}=\{\mathcal{E}_1,q{\cdot}s_2,\sqrt{-q^2}p_1{\cdot}s_2,\mathcal{E}_2,q{\cdot}s_1,\sqrt{-q^2}p_2{\cdot}s_1 \}$, for the \textit{regular} operators, and the spin operator basis $\mathcal{H}\to\tilde{\mathcal{H}}=\mathcal{H} \cup\{|q||s_1|,|q||s_2| \}$ for the \textit{exotic} terms appearing starting at the fifth order in spin, 
we write the 2PM contribution of the triangle cut  in the form
\begin{equation}\label{eq:cons}
M_{\text{bare}}^{\text{2PM}}=\frac{\pi G^2}{\sqrt{-q^2}}\sum_{i,j}\Big( A_{i,j}\mathcal{H}^{\otimes^i}_j + B_{i,j}\tilde{\mathcal{H}}^{\otimes^i}_j \Big)\,.
\end{equation}
Then, the main outputs from the LS evaluation are the  $\{A_{i,j},B_{i,j}\}$ coefficients for a given  spin structure in $\{\mathcal{H}^{\otimes^i}_j,\tilde{\mathcal{H}}^{\otimes^i}_j \}$ respectively. 

Up to the fourth order in spin, the LS construction easily recovers the triple cut coefficients  reported in \cite{Bern:2020buy,Kosmopoulos:2021zoq,Chen:2021kxt}  for SE Kerr BHs with generic spin orientations, upon evaluating to zero the contact deformations present at spin 4 in \eqref{eq:ansatzspin}, as dictated by the Teukolsky solution (\cref{tab:Teukolskysolutions}).
In this work, we extend the 2PM computation up to the sixth order in  spin including both, \textit{regular} and \textit{exotic} contributions to  \eqref{eq:cons}. We provide  LS results for  generic coefficients parametrizing the Compton ansatz \eqref{eq:ansatzspin}, which  can  then be specialized   (if desired) to Teukolsky solutions (\cref{tab:Teukolskysolutions}) for the  matching prescription described above.

Let us for brevity include here only the explicit 
  results for the $s_1^5\times s_2^0$ sector of the 2PM triangle, leaving the results for the additional spin sectors, and up to the sixth order,  
 for the ancillary files for this work \cite{ancill}. 
 The spin structures for this sector are  summarized  in \cref{tab:spin5}, and  the  explicit coefficients for generic contact deformations are   given   in \cref{tab:spin5gen}. Additionally, we include in \cref{tab:dictionary} a dictionary that maps  between our generic coefficient amplitude  and the one reported in \cite{Bern:2022kto}, hence providing a  connection of the \textit{regular} operators in \eqref{eq:ansatzspin} to the Lagrangian construction in \cite{Bern:2022kto}. 
 
Before moving to studding the content of these tables in more detail, and since 
 most  of the results of this work are included as  ancillary files \cite{ancill}, let us here summarise the content of such files. We present three files named \texttt{CoefficientsFile.wl, 2PMAmplitudeEikonalScatteringAngle.wl} and \texttt{CanonicalObservables.wl}. The first of them contains the list of coefficients for the 1PM and 2PM amplitude  as appearing in \eqref{eq:cons}. In addition, the coefficients for the 1PM and 2PM eikonal phase for both CovSSC and CanSSC (see \eqref{eq:1pmeik} and \eqref{eq:eik}) are included. Finally, replacement rules for the free coefficients of the Compton amplitude \eqref{eq:ansatzspin}, as coming from Teukolsky,  or the   spin-shift-symmetry solutions (see \cref{tab:Teukolskysolutions}), and 
the  dictionary included in \cref{tab:dictionary} are provided.  The second file contains an explicit implementation of eqs. \eqref{eq:cons} and \eqref{eq:eik}, as well as the explicit results for the aligned spin scattering angle (see below)  in the CovSSC up to six order in spin. The remaining file, contains the results for 2PM canonical impulse and spin-kick (see  \eqref{eq:master_observable}), up to sixth order in spin, including contributions from both, \textit{regular} and \textit{exotic} spin operators included in  the Compton amplitude. Let us recall  contributions from \textit{exotic} operators appear only when at least
one of the two spins is beyond the fourth order.

\begin{table}
\begin{centering}
\setlength{\tabcolsep}{1.0pt} 
\setlength\extrarowheight{2.0pt} 
\begin{tabular}{c|c|c|c|c|c|c}
 & $j$ &  & $j$ &  & $j$ & \tabularnewline
\hline 
\hline 
\multirow{2}{*}{$\frac{\mathcal{H}_{j}^{\otimes5}}{\mathcal{E}_{1}}$} & $1$ & $\mathcal{E}_{1}^{4}$ & $2$ & $-q^{2}(p_{2}\cdot s_{1})^{2}\mathcal{E}_{1}^{2}$ & $3$ & $q^{4}(p_{2}\cdot s_{1})^{4}$\tabularnewline
 & $4$ & $(\ensuremath{q\cdot s_{1})^{2}\mathcal{E}_{1}^{2}}$ & $5$ & $-q^{2}(q\cdot s_{1})^{2}(p_{2}\cdot s_{1})^{2}$ & $6$ & $(q\cdot s_{1})^{4}$\tabularnewline
\hline 
$\frac{\mathcal{\tilde{H}}_{j}^{\otimes5}}{(p_2\cdot s_1)\mathcal{E}_{1}|s_{1}|}$ & $1$ & $-q^{2}\mathcal{E}_{1}^{2}$ & $2$ & $q^{4}(p_{2}\cdot s_{1})^{2}$ & $3$ & $-q^{2}(q\cdot s_{1})^{2}$\tabularnewline
\end{tabular}
\caption{The independent spin structures for the 
$s_{1}^{5}\times s_{2}^{0}$ sector of the 2PM amplitude \eqref{eq:cons}.}
\label{tab:spin5}
\par\end{centering}
\end{table}

\begin{table*}[ht]
\begin{centering}
\setlength\extrarowheight{8pt}
\resizebox{\textwidth}{!}{
\begin{tabular}{c|c|c|c|c|c}
 & $j$ & \multicolumn{1}{c}{} &  & $j$ & \tabularnewline
\cline{1-5} \cline{2-5} \cline{3-5} \cline{4-5} \cline{5-5} 
\multirow{6}{*}{$A_{5,j}$} & $1$ & \multicolumn{2}{c|}{$\frac{i\sigma\left(7-13\sigma^{2}\right)}{48\left(-1+\sigma^{2}\right)^{3}m_{1}^{7}m_{2}^{3}}-\frac{i\sigma\left(-96-60c_2^{(1)}+480c_2^{(2)}+135c_3^{(0)}+7\sigma^{4}\left(-4+120c_2^{(2)}+15c_3^{(0)}-15c_3^{(1)}\right)-90c_3^{(1)}+\sigma^{2}\left(304+60c_2^{(1)}-1320c_2^{(2)}-240c_3^{(0)}+195c_3^{(1)}\right)\right)}{1920\left(-1+\sigma^{2}\right)^{3}m_{1}^{8}m_{2}^{2}}$} & \multirow{3}{*}{} & \tabularnewline
 & $2$ & \multicolumn{2}{c|}{$\frac{i\sigma\left(-5+11\sigma^{2}\right)}{24\left(-1+\sigma^{2}\right)^{3}m_{1}^{5}m_{2}^{3}}-\frac{i\sigma\left(4\left(-36+75c_2^{(1)}+30c_2^{(2)}-\sigma^{2}\left(82+180c_2^{(1)}+240c_2^{(2)}-195c_3^{(0)}\right)+7\sigma^{4}\left(4+15c_2^{(1)}+30c_2^{(2)}-15c_3^{(0)}\right)-90c_3^{(0)}\right)+15\left(9-16\sigma^{2}+7\sigma^{4}\right)c_3^{(1)}\right)}{1920\left(-1+\sigma^{2}\right)^{3}m_{1}^{6}m_{2}^{2}}$} &  & \tabularnewline
 & $3$ & \multicolumn{2}{c|}{$\frac{i\sigma\left(-1+3\sigma^{2}\right)}{48\left(-1+\sigma^{2}\right)^{3}m_{1}^{3}m_{2}^{3}}-\frac{i\sigma\left(3\left(-12+40c_2^{(1)}-40c_2^{(2)}-55c_3^{(0)}+25c_3^{(1)}\right)+\sigma^{2}\left(16+20\left(-13+7\sigma^{2}\right)c_2^{(1)}+120c_2^{(2)}+5\left(68-35\sigma^{2}\right)c_3^{(0)}+5\left(-29+14\sigma^{2}\right)c_3^{(1)}\right)\right)}{640\left(-1+\sigma^{2}\right)^{3}m_{1}^{4}m_{2}^{2}}$} &  & \tabularnewline
\cline{4-6} \cline{5-6} \cline{6-6} 
 & $4$ & $-\frac{i\sigma\left(14\sigma^{2}\left(-4+120c_2^{(2)}+15c_3^{(0)}-15c_3^{(1)}\right)+3\left(16+20c_2^{(1)}-280c_2^{(2)}-60c_3^{(0)}+45c_3^{(1)}\right)\right)}{1920\left(-1+\sigma^{2}\right)m_{1}^{6}}$ & $\frac{i\left(6c_4^{(0)}+7\left(-1+\sigma^{2}\right)c_4^{(1)}+2\left(4-7\sigma^{2}\right)c_4^{(2)}\right)}{64\left(-1+\sigma^{2}\right)m_{1}^{6}m_{2}}$ & $1$ & \multirow{3}{*}{$B_{5,j}$}\tabularnewline
 & $5$ & $\frac{30i\sigma\left(4+12c_2^{(1)}-15c_3^{(0)}+6c_3^{(1)}\right)-7i\sigma^{3}\left(16+60c_2^{(1)}+120c_2^{(2)}-60c_3^{(0)}+15c_3^{(1)}\right)}{1920\left(-1+\sigma^{2}\right)m_{1}^{4}}$ & $\frac{i\left(2\left(-4+7\sigma^{2}\right)c_4^{(0)}-7\left(-1+\sigma^{2}\right)c_4^{(1)}-6c_4^{(2)}\right)}{64\left(-1+\sigma^{2}\right)m_{1}^{4}m_{2}}$ & $2$ & \tabularnewline
 & $6$ & $-\frac{i\sigma\left(-3+7\sigma^{2}\right)m_{2}^{2}\left(-4+120c_2^{(2)}+15c_3^{(0)}-15c_3^{(1)}\right)}{1920m_{1}^{4}}$ & $\frac{i\left(-1+7\sigma^{2}\right)m_{2}\left(c_4^{(1)}-2c_4^{(2)}\right)}{64m_{1}^{4}}$ & $3$ & \tabularnewline
\hline 
\end{tabular}}
\par\end{centering}
\vspace{-0.2cm}
\caption{2PM amplitude coefficients  for the $s_1^5\times s_2^0$ sector of the two-body problem \eqref{eq:cons}.  }
\label{tab:spin5gen}
\end{table*}

\begin{table}

\begingroup\makeatletter\def\f@size{5.9}\check@mathfonts
\def\maketag@@@#1{\hbox{\m@th\large\normalfont#1}}%

\begin{equation*}
\boxed{
\begin{split}
c_2^{(2)}&\to\frac{374-165E_{1}+513E_{3}-33E_{4}-330E_{5}-495c_3^{(0)}+495c_3^{(1)}}{3960}\,,\\
c_2^{(1)}&\to\frac{153E_{3}-11\left(4+3E_{4}+15E_{5}\right)+2475c_3^{(0)}-990c_3^{(1)}}{1980}\,,\\
E_{7}&\to\frac{17}{30}+\frac{21E_{3}}{22}-\frac{E_{4}}{2}-\frac{E_{5}}{2}\,,\,\,E_{2}\to-E_{1}+\frac{3}{2}\left(E_{3}+E_{4}\right)\,,\,\,H_3\to\frac{3}{2}\,, \\
C_{i}&\to1\,,i=2,3,4,5\,,\,\, H_2\to1\,,\,\,   E_{6}\to\frac{1}{6}\left(-2-3E_{3}\right)\,,\,\,H_{5}\to\frac{15E_{3}}{11}\,.
\end{split}
}
\end{equation*}
\endgroup
\vspace{-0.5cm}
\caption{Coefficient dictionary relating the \textit{regular} terms of the  $s_1^5\times s_2^0$ sector of the  2PM amplitude in this letter and the generic 2PM amplitude included in the ancillary files for ref. \cite{Bern:2022kto}. The fixing $C_i,H_2\to1$ agrees with the   minimal coupling matching of  the 3-pt amplitude to the linearized  Kerr metric, built in  the Compton ansatz  \eqref{eq:ansatzspin}. Spin-shift symmetric solutions  evaluated using the values of the last two columns of \cref{tab:Teukolskysolutions} agree with those given in  \cite{Bern:2022kto}.  }
\label{tab:dictionary}
\end{table}

\begin{table}
\centering{}%
\setlength{\tabcolsep}{3.8pt} 
\setlength\extrarowheight{4.0pt}
\begin{tabular}{c|c|c|c|c}
 & $j$ &  & $j$ & \tabularnewline
\hline 
\multirow{3}{*}{$A_{5,j}$} & $1$ & $\frac{i\sigma\left(14m_{1}-2(13m_{1}+8m_{2})\sigma^{2}+7m_{2}\sigma^{4}\right)}{96m_{1}^{8}m_{2}^{3}(\sigma^{2}-1)^{3}}$ & $4$ & $\frac{7i\sigma^{3}}{48m_{1}^{6}(\sigma^{2}-1)}$\tabularnewline
 & $2$ & $\frac{i\sigma\left(2m_{1}(11\sigma^{2}-5)+m_{2}(24-43\sigma^{2}+28\sigma^{4})\right)}{48m_{1}^{6}m_{2}^{3}(\sigma^{2}-1)^{3}}$ & $5$ & $\frac{7i\sigma(4\sigma^{2}-3)}{48m_{1}^{4}(\sigma^{2}-1)}$\tabularnewline
 & $3$ & $\frac{i\sigma\left(m_{1}(6\sigma^{2}-2)+m_{2}(51-104\sigma^{2}+56\sigma^{4})\right)}{96m_{1}^{4}m_{2}^{3}(\sigma^{2}-1)^{3}}$ & $6$ & $\frac{im_{2}^{2}\sigma(7\sigma^{2}-3)}{96m_{1}^{4}}$\tabularnewline
\hline 
\multirow{2}{*}{$B_{5,j}$} & $1$ & $\frac{i\eta(2+7\sigma^{2})}{24m_{1}^{6}m_{2}(\sigma^{2}-1)}$ & $3$ & $\frac{i\eta m_{2}(-1+7\sigma^{2})}{24m_{1}^{4}(\sigma^{2}-1)}$\tabularnewline
 & $2$ & $\frac{i\eta(-5+14\sigma^{2})}{24m_{1}^{4}m_{2}(\sigma^{2}-1)}$ &  & \tabularnewline
\end{tabular}
\vspace{-0.1cm}
\caption{2PM LS  coefficients for the  $s_1^5\times s_2^0$ sector of the 2PM amplitude explicitly evaluated on the
Teukolsky solutions \cref{tab:Teukolskysolutions} for $\alpha=1$.}
\label{tab:amp5}
\end{table}

\sectionskip
\Section{Spin-shift symmetry violation for Teukolsky solutions and the high-energy limit.} 
It is illustrative to consider explicitly the   2PM LS coefficients evaluated on the Teukolsky solutions given in \cref{tab:Teukolskysolutions}; we include them in \cref{tab:amp5}.
Inspection of these results  reveals a breaking of  the   invariance of the amplitude   under the transformation   $
 a_i^\mu\to a_i^\mu+\varsigma_i q^\mu/q^2  \,, 
$ with  $\varsigma_i$  arbitrary constants, observed for the $a_i^{n\le4}$ cases \cite{Bern:2022kto,Aoude:2022trd,Aoude:2022thd}. This happens  not only due to the presence of the  \textit{exotic} operators but also because of the non-zero contributions  of  \textit{regular} operators of the form $q\cdot s_i$ in \cref{tab:spin5}. One can also check  shift-symmetric solutions for the Compton amplitude (\cref{tab:Teukolskysolutions} ) induce a shift symmetric 2PM amplitude, as first observed in \cite{Aoude:2022thd}.

Let us now comment on the high energy behavior of the 2PM amplitude for Teukolsky solutions  (\cref{tab:amp5}). In the high energy limit ($\sigma\to\infty$), the 1PM amplitude scales  as $\mathcal{O}(\sigma^2)$  (here $\mathcal{E}_i\sim \sigma$). Having a well-defined 2PM amplitude in the high energy limit means it should  grow no faster than the tree amplitude,   as $\sigma\to \infty$ \cite{Bern:2022kto}. This is however not the case for Teukolsky solutions of \cref{tab:amp5}. For instance, the term $A_{5,1}\mathcal{E}_i^5$ in \eqref{eq:cons} grows as $\sim \sigma^4$ as $\sigma\to\infty$, and analogously for the remaining contributions. This  singular high-energy behavior propagates  to the  2PM observables as we will discuss  below in the context of the  aligned spin scattering angle. 

\sectionskip
\Section{Eikonal phase and the aligned spin limit.}
The LS computation provides us with   the  ingredients needed to obtain  the 2PM  eikonal phase \eqref{eq:eikonalgen} for spinning objects satisfying  both CovSSC and CanSSC, the latter obtained from the former by the shift of  the impact parameter  $\boldsymbol{b}\to \boldsymbol{b}+\sum_{i=1,2}\left(\frac{-\tau}{E_i+m_i}\right) \frac{\boldsymbol{p}\times \boldsymbol{s}_i}{m_i}$, a consequence of dressing the amplitude with the   rotation factors \eqref{eq:1PMdressed}. Results for the eikonal phase in the CanSSC will be needed when evaluating canonical observables via \eqref{eq:master_observable}. We evaluate explicitly \eqref{eq:eikonalgen} and present the results for the eikonal phase in the CovSSC and in CoM coordinates in the ancillary files \cite{ancill}. The eikonal phase is  organized schematically as 

\begin{equation} \label{eq:eik}
\chi_{\tau=0}^{(\text{2PM})}  =  \sum_{i,j}\frac{\pi G^2}{ b^{2j+1}}\Big( L_{i,j}\, b\cdot \mathcal{S}^j+Ld_{i,j} \,b\cdot \tilde{\mathcal{S}}^j\Big) \,,
\end{equation}
with $\{\mathcal{S}^j,\tilde{\mathcal{S}}\}$   operators in the CoM version of $\{(\{b\} \cup\mathcal{H})^{\otimes j},(\{b\} \cup\tilde{\mathcal{H}})^{\otimes j}\}$ respectively,  $b^\mu=(0,\boldsymbol{b},0)$, and the coefficients $\{L_{i,j},Ld_{i,j}\}$ are functions of $m_i,E_i,\sigma$, and linear combinations of the amplitude coefficients $\{A_{i,j},B_{i,j} \}$ respectively.   From this, we then specialize the eikonal phase to  the case in which the  rotating objects have their spins aligned in the direction of the angular momentum of the system, $\boldsymbol{b}\cdot\boldsymbol{s}_i=\boldsymbol{p}\cdot\boldsymbol{s}_i=0$ (see $e.g.$ \cite{Kosmopoulos:2021zoq})  and compute the 2PM aligned spin scattering angle via $ \frac{\partial \chi_{\tau=0}}{\partial |b|} $. In this limit, the \textit{exotic} operators do not contribute to the scattering angle as  observed in \cite{Bautista:2022wjf}, therefore for Teukolsky solutions,  the angle itself is independent of the analytic continuation procedure used to match \eqref{eq:ansatzspin} to the GW scattering process. Results up to the sixth order in spin for the contributing \textit{regular} operators are  included in the ancillary files for generic contact deformations in \eqref{eq:ansatzspin} \cite{ancill}.

Let us now comment on the comparison of the aligned spin scattering angle  results in this work to the ones  presented in \cite{Guevara:2018wpp,Bautista:2022wjf} for generic, spin-shift-symmetric \cite{Bern:2022kto,Aoude:2022trd,Aoude:2022thd}, and Teukolsky Compton  coefficients \cite{Bautista:2022wjf}. 
The  HCL prescription  of  \cite{Guevara:2018wpp}  to compute the aligned spin scattering angle  by fixing  $\beta=1$, in turn hiddenly sets $q{\cdot}s_j\to i \frac{\mathcal{E}_j}{m_1 m_2\sigma v}$  as well as $|q|\to0 \Rightarrow |q|p_i{\cdot} s_j\to0$   in  \eqref{eq:spin_op}, as can be seen from the last line of \eqref{eq:kinematics HCL}. The latter replacement does not have any problem since in the aligned spin limit  the equality    $\boldsymbol{p}{\cdot}\boldsymbol{s}_i=0$ is satisfied. Similarly, the  former  identification does not possess any subtleties  for combinations of spin operators of the form  $(k_2-k_3){\cdot} s_i$, since these combinations themselves are independent of $q^\mu$, as can be checked by direct inspection of  eq. \eqref{eq:spin_op}. However, for spin operators where  $k_i{\cdot} s_i$ have individual appearances -- as  is the case for some of the contact deformations in \eqref{eq:ansatzspin} -- the identification $q.s_j\to i \frac{\mathcal{E}_j}{m_1 m_2\sigma v}$ discards $\mathcal{O}(q^2)$ terms (see \eqref{eq:kinematics HCL}) that are important for the aligned spin angle.
For quadratic terms, for instance, this map is equivalent to setting
$
(q{\cdot}s_i)^2\to(q{\cdot}s_i)^2 - q^2 s_i^2\,,
$
which in  two-dimensional impact parameter space  implies
$
3 (\boldsymbol{b}{\cdot}\boldsymbol{s}_i)^2-\boldsymbol{b}^2\boldsymbol{s}_{\perp i}^2\to3 (\boldsymbol{b}{\cdot}\boldsymbol{s}_i)^2-2\boldsymbol{b}^2\boldsymbol{s}_{\perp i}^2 \,, 
$ -- with $\boldsymbol{s}_{\perp i}$ the components of the spin along  $\boldsymbol{b}$ --
therefore removing some terms that survive in the aligned spin limit ($\boldsymbol{b} {\cdot}\boldsymbol{s}_i=0$). An analogous analysis  follows for higher spin contributions.  Interestingly, this identification  removes the terms in the amplitude that did not have a  well-defined high energy limit.  The spin-shift symmetric result of \cite{Bern:2020buy,Aoude:2022thd}  has also a well-defined high-energy limit. 
Finally, let us note the angle for the lower spin cases \cite{Guevara:2018wpp} did not face this ambiguity since only the combination $(2w-k_2-k_3)\cdot a_i$ -- independent of $q{\cdot}a_i$ terms --  appeared in the Compton amplitude.  

Keeping all the contributions to the scattering angle, and continuing with  the theme of the spin 5  sector  for  briefness, the aligned spin angle takes the form
\be
\begin{split}
\theta^{(5)} =&\,\theta^{(5)}_{\text{BGKV}}-  \frac{9\pi E G^2}{256\,b^7\, v(1-v^2)} \Big[(a_1^5 m_2{+}a_2^5 m_1)K_1\\
&\hspace{2.5cm} +(a_1^3 m_2{+}a_2^3 m_1)a_1a_2\,K_2 \Big]\,, 
\end{split}
\ee
where  $K_1 = 80(1-v^2)c_{2}^{(1)}+40(8+13v^2)c_{2}^{(2)}+(4-11v^2)(4-15c_{3}^{(0)})-105v^2c_{3}^{(1)}$ and $K_2=-40\big[2(1-v^2)c_{1}^{(1)}+(8+13v^2)c_{1}^{(2)} \big]$, and with 
$\theta^{(5)}_{\text{BGKV}}$   the scattering angle reported in the arXiv v2 of \cite{Bautista:2022wjf}.  Teukolsky solutions (\cref{tab:Teukolskysolutions}) give   $K_1|_{\text{Teuk.}} = 20(12-5v^2)$ and $K_2|_{\text{Teuk.}}=0$, therefore resulting into a scattering angle divergent as $v\to1$ as expected from the discussion above. Let us stress this result is independent of the analytic continuation procedure for  the matching of \cref{eq:ansatzspin} to the Teukolsky solutions. Finally,   for a shift-symmetric amplitude (see the last two columns of \cref{tab:Teukolskysolutions}), $K_1=K_2=0$ recovers the result of \cite{Bern:2022kto,Aoude:2022thd}, and the angle is well behaved in the high-energy limit. The  sixth order in spin angle can be obtained  analogously, explicit results  can be found  in  the ancillary files \cite{ancill}. Up to $\mathcal{O}(a^6)$, and in the  probe limit, our   results  are in complete  agreement  with the   ones reported in \cite{Damgaard:2022jem}; contact deformations of the Compton amplitude do not contribute to the 2PM  angle in this case. In addition, only the next  order in the symmetric mass ratio  is needed to fully obtain  our 2PM results, as expected from the spin version of the   mass polynomiality  rule \cite{Damour:2019lcq}.

\sectionskip
\Section{2PM canonical observables.} 
We are finally in ready to compute the \textit{ conservative} 2PM canonical observables for  $2\to2$ scattering of SE Kerr BHs. For this, we will follow the prescription provided by the authors of  \cite{Bern:2020buy,Kosmopoulos:2021zoq}. 
It was noticed in these references that  the   conservative  observables $\Delta\mathcal{O}\in\{\Delta\boldsymbol{p}_\perp,\Delta\boldsymbol{s}_a \}$ at 2PM   can be  obtained from the eikonal phase in the CanSSC $\chi\equiv \chi_{\tau=1}=\chi^{\text{1PM}}_{\tau=1}+\chi^{\text{2PM}}_{\tau=1}+\cdots$,  via
\be\label{eq:master_observable}
\begin{split}
\Delta \mathcal{O} =&
 -\{\mathcal{O},\chi\} -\frac{1}{2}\{\chi,\{\mathcal{O},\chi\} \}-\mathcal{D}_{SL}(\chi,\{\mathcal{O},\chi\}) \\& +\frac{1}{2}\{\mathcal{O},\mathcal{D}_{SL}(\chi,\chi)\}\,,
\end{split}
\ee 
with the Poisson bracket given by
\be
\{f,g\}{=} \frac{\partial f}{\partial p_\perp^j} \frac{\partial g}{\partial b^j}{-}\frac{\partial g}{\partial p_\perp^j} \frac{\partial f}{\partial b^j} {+}\sum_{a=1,2}\epsilon^{ijk} \frac{\partial f}{\partial s^i_a}\frac{\partial g}{\partial s^j_a}s_{a,k}\,,
\ee
and the spin derivative operator
\be
\mathcal{D}_{SL} (f,g){=} \frac{1}{\boldsymbol{p}^2}\sum_{a,1,2}\left(\frac{\partial f}{\partial s_a^j}\frac{\partial g}{\partial b^j} \boldsymbol{s}_a{\cdot}\boldsymbol{p} {-}p^j \frac{\partial f}{\partial s_a^j} \boldsymbol{s}_a{\cdot} \nabla_{\boldsymbol{b}}g \right)
\ee
Using this prescription, and the results for the 1 and 2PM eikonal phase derived above,  we have computed the 2PM  observables up to the six order in spin for generic contact deformations in \cref{eq:ansatzspin}, which can then specialize to Teukolsky solutions. The expressions are however  too long to be included in this letter and we, therefore, provide them in the  ancillary material for this work \cite{ancill}.  We include results for the transverse impulse and individual spin kicks in the CanSSC up to the sixth order in spin for  both  \textit{regular } and \textit{exotic} contributions to the 2PM amplitude   in the CoM frame. Up to the fourth order in spin our results completely agree with those reported in \cite{Bern:2020buy,Kosmopoulos:2021zoq,Chen:2021kxt} upon setting to zero the contact deformations at this order, and after the authors of \cite{Chen:2021kxt} fixed some of the reported observables that had an initial disagreement with the ones presented in this work.

\sectionskip
\Section{Conclusions.} 
In this work, we have computed the  canonical observables for the \textit{conservative} SE Kerr   two-body problem at second order in the PM expansion and up to sixth order in spin,  for generic spin orientation. The results  in this work are presented for generic Compton contact deformations, which can be specialized to Teukolsky solutions. In the latter case, the 2PM amplitude  breaks the conjecture spin-shift symmetry for Kerr BHs \cite{Bern:2022kto,Aoude:2022trd,Aoude:2022thd},
  however  producing observables  with a non-smooth high energy behavior. This leaves as an open problem  understanding if the  unhealthy high-energy behavior  is a consequence of the analytic continuation to the SE Kerr region and 
 observables for actual Kerr  solutions  ($a^\star\le1$) feature a well-defined  high energy limit \footnote{Non-smooth high energy limit in the two-body context  has been recently reported in the 4PM results for scalar BHs \cite{Dlapa:2022lmu}, which are expected to 
 be improved by the  non-perturbative
solutions  \cite{Gruzinov:2014moa,DiVecchia:2022nna}.
 }, perhaps guiding    possible realizations of the spin-shift symmetry in this scenario.
For the generic coefficient case, we have presented a dictionary that maps the Compton operators in this letter to the Lagrangian and Hamiltonian constructions in \cite{Bern:2022kto}. Finally, the identification of \textit{regular} and \textit{exotic} contributions with \textit{true conservative} and \textit{absorptive} contributions in the Kerr binary problem is a subject that requires further scrutiny and we leave for future work. 

\sectionskip
\Section{Acknowledgments.}   
We would like to thank  Rafael Aoude, Stefano De Angelis,  Leonardo de la Cruz, Alfredo Guevara, Kays Haddad, Carlo Heissenberg, Andreas Helset, Chris Kavanagh, Dimitrios Kosmopoulos, David Kosower, Andres Luna and   M. V. S. Saketh for  useful discussion. We are also grateful to  Jung-Wook Kim and  Ming-Zhi Chung for the continuum discussion and for sharing an updated version of their 2PM observables \cite{Chen:2021kxt} up to spin 4. We would like  to especially thank Justin Vines for sharing his unpublished notes on the generalization of the HCL for non-aligned spins, and Leonardo de la Cruz for his comments on the manuscript. 
This work has been supported by the European Research Council under Advanced Investigator Grant ERC–AdG–885414.

\bibliography{references}

\newpage
\onecolumngrid
\newpage
\appendix
\section{Appendix A: Triangle Leading singularity and the holomorphic classical limit for generic spin orientation }\label{app_A}

In this appendix, we present an extension of  the Holomorphic Classical Limit (HCL)  \cite{Guevara:2017csg} and the  triangle  Leading Singularity (LS) \cite{Cachazo:2017jef} constructions  for  spinning particles for generic  spin directions.  This is based on an unpublished note by Justin Vines to whom we are grateful.

We start by constructing a suitable local 4-dimensional reference frame \textit{all'a Penrose, Rindler, and Chandrasekhar }\cite{Penrose1984}  as follows: Given a two-dimensional basis of massless  spinors $\mathcal{L}= \{|\chi\rangle ,|\psi \rangle  \}$, together with  the co-basis  $\tilde{\mathcal{L}}=\{[\chi| ,[\psi| \}$, and whose   elements  satisfy the normalization conditions
$
\langle \chi\psi\rangle=[\chi\psi]=\frac{1}{\sqrt{2}},
$
we can adapt the spinor bases to the vector tetrad  $\{e^{\mu}_{(a)} \}=\{ \frak{t}^\mu,\frak{x}^\mu,\frak{y}^\mu,\frak{z}^\mu\}$,  with $\frak{t}^2=-1$ and $\frak{x}^2=\frak{y}^2=\frak{z}^2=-1$,  via the usual vector-spinors map  $V^{\alpha\dot{\alpha}}=(\sigma^\mu)^{\alpha\dot{\alpha}}v_\mu$, as follows:
\begin{equation}
\begin{split}
\frak{T}^{\alpha\dot{\alpha}}&=(\sigma^\mu)^{\alpha\dot{\alpha}}\frak{t}_\mu =\sqrt{2}(|\chi\rangle[\psi|+|\psi\rangle[\chi|)\,,\\
\frak{Z}^{\alpha\dot{\alpha}}&=(\sigma^\mu)^{\alpha\dot{\alpha}}\frak{z}_\mu =\sqrt{2}(|\chi\rangle[\psi|-|\psi\rangle[\chi|)\,,\\
\frak{X}^{\alpha\dot{\alpha}}&=(\sigma^\mu)^{\alpha\dot{\alpha}}\frak{x}_\mu =\sqrt{2}(|\chi\rangle[\chi|+|\psi\rangle[\psi|)\,,\\
\frak{Y}^{\alpha\dot{\alpha}}&=(\sigma^\mu)^{\alpha\dot{\alpha}}\frak{y}_\mu =i\sqrt{2}(|\chi\rangle[\chi|-|\psi\rangle[\psi|)\,.
\end{split}
\end{equation}
Here $\sigma^\mu$ are the Pauli matrices and shall not  be confused with the Lorentz boost factor $\sigma$ which is a scalar. 

This tetrad spans the  4-dimensional flat spacetime and therefore any vector (including the spin) can be constructed from a linear combination of vectors in $\{e^\mu_{(a)}\}$.  
Notice  in general the spinors in $\mathcal{L}$ are not related to those in $\tilde{\mathcal{L}}$ by complex conjugation and therefore the tetrad decomposition admits vectors with  complex values. Thus, a generic vector $q^\mu$  will be given by  the decomposition
\be\label{eq:mom-transfer}
Q^{\alpha\dot\alpha}=(\sigma^\mu)^{\alpha\dot{\alpha}}q_\mu= a_q |\chi\rangle [\chi|+b_q |\chi\rangle[ \psi|+c_q|\psi \rangle[ \chi|+d_q |\psi \rangle [\psi|\,,
\ee
where $a_q= \frac{1}{2}\langle \psi| Q|\psi]$, and analogously for the other components. 
Lorentz invariant products  are  obtained in the usual form
\begin{align}
q^2&=\frac{1}{2}\text{tr}(Q\cdot\bar{Q})=\frac{1}{2}(c_q b_q-a_q d_q)\,,\,\,\,\,
p{\cdot}q=\frac{1}{2}\text{tr}(Q\cdot\bar{P})=\frac{1}{4}(b_q c_p+b_p c_q -a_q d_p -a_p d_q)\,.
\end{align}

\textit{2PM Triangle kinematics}: 
The next task is to use this reference frame to parametrize the momenta of the particles  in the triangle cut \cref{leading-singularity}. Unlike for the aligned spin scenario, here we will take  the momentum transfer  $q=k_2+k_3$ so that   $q^2=2k_2{\cdot}k_3\ne0$. We orient the tetrad $\{e_{(a)}^\mu\}$ in such a way that the  BH 2 is at rest, the  BH 1 moves in the $\frak{z}$-direction with relative velocity $v$, and the component of $q^\mu$ orthogonal to $p_1^\mu$ and $p_2^\mu$ lies  in the $\frak{x}$-direction. It follows then 
\be\label{eq:external_tetrad}
p_1^\mu = m_1 \sigma(\frak{t}^\mu+v\frak{z}^\mu)\,,\,\,\,p_2^\mu = m_2 \frak{t}^\mu\,,\,\,\,q{\cdot}\frak{y}=0\,.
\ee

The momentum transfer $q^\mu$ can be parametrized using the decomposition \eqref{eq:mom-transfer}. With help of  the on-shell conditions $p_1{\cdot}q=q^2/2,\,\,p_2{\cdot}q=-q^2/2$ and $q{\cdot}\frak{y}=0$, and solving for  the coefficients $a_q,b_q,c_q,d_2$ in terms of the kinematic variables $m_1,m_2,\sigma,|q|$, where $|q|=\sqrt{-q^2}$, one explicitly gets
\be\label{eq:onshell_sol}
b_q=\frac{m_2+m_1(1+v)\sigma}{\sqrt{2}m_1 m_2 \sigma v}|q|^2\,,\,\,\,c_q= - \frac{m_2+m_1(1-v)\sigma}{\sqrt{2}m_1 m_2 \sigma v}|q|^2\,,\,\,
d_q=a_q=\pm\sqrt{b_q^2c_q^2+2|q|^2} \,.
\ee

\textit{HCL parametrization}:
The final task is to find a suitable parametrization for the internal massless momenta in such a way the classical limit of the triangle diagram \cref{leading-singularity} is easily obtained. For that, let us  introduce the new spinor bases $\{|\hat{\lambda}\rangle,|\hat{\eta}\rangle\}$ and $\{[\hat{\lambda}|,[\hat{\eta}|\}$ for the first massive line, together with   $\{|\lambda\rangle,|\eta\rangle\}$ and $\{[\lambda|,[\eta|\}$ for the second massive line, following Guevara \cite{Guevara:2017csg}. In these new bases, the external momenta are parametrized as
\begin{equation}\label{eq:kinematics HCL}
    \begin{split}
     P_1 &=|\hat{\eta}]\bra{\hat{\lambda}}+|\hat{\lambda}]\bra{\hat{\eta}}\,,\,\,\,\,
     \tilde{P}_1 =  \beta^\prime|\hat{\eta}]\bra{\hat{\lambda}}+\frac{1}{\beta ^\prime}|\hat{\lambda}]\bra{\hat{\eta}} +|\hat{\lambda}]\bra{\hat{\lambda}},\\
     P_2 &=|\eta]\bra{\lambda}+|\lambda]\bra{\eta}\,,\,\,\,\,
     \tilde{P}_2 =  \beta|\eta]\bra{\lambda}+\frac{1}{\beta}|\lambda]\bra{\eta} +|\lambda]\bra{\lambda},\\
     Q &=  P_1-\tilde{P}_1= -P_2+\tilde{P}_2\,,\,\,\,\,
     \Rightarrow \,\,|q|^2=m_2^2\frac{(\beta-1)^2}{\beta}\,,
    \end{split}
\end{equation}
where $Q$ is the complex momentum transfer matrix. Thus, as $|q|\to 0$ (as $\beta\to1$) one  recovers the usual HCL result $Q\to |\lambda]\langle \lambda|$, but now we keep subleading terms in $(\beta-1)$. 
The on-shell conditions $P_1^2=\tilde{P}_1^2=m_1^2$ and $P_2^2=\tilde{P}_2^2=m_2^2$, impose the normalization for the spinors $\braket{\hat{\lambda}\hat{\eta}}=[\hat{\lambda}\hat{\eta}]=m_1$ and $\braket{\lambda\eta}=[\lambda\eta]=m_2$. 
For the internal gravitons, the spinor helicity variables are analogously parametrized via
\begin{equation}\label{eq:kinematics HCL gravitons}
    \begin{split}
        \ket{k_2} &= \frac{1}{\beta+1}\left(\left(\beta^2-1 \right)\ket{\eta} -\frac{1+\beta y}{y}\ket{\lambda}\right),\,\,\,|k_2] = \frac{1}{\beta+1}\left(\left(\beta^2-1\right)y|\eta]+(1+\beta y)|\lambda]\right),\\
        \ket{k_3} &= \frac{1}{\beta+1}\left(\frac{\beta^2-1}{\beta} \ket{\eta} +\frac{1-y}{y}\ket{\lambda}\right),\,\,\,|k_3] = \frac{1}{\beta+1}\left(-\beta\left(\beta^2-1\right)y|\eta]+(1-\beta^2 y)|\lambda]\right).
    \end{split}
\end{equation}
Here $y$ is the loop integration parameter entering  in \eqref{LS}. 
Imposing consistency between the HCL \eqref{eq:kinematics HCL} and the tetrad \eqref{eq:mom-transfer} \eqref{eq:external_tetrad}  parametrizations, allows us   to solve for $\{|\lambda\rangle,|\eta\rangle,[\lambda|,[\eta| \}$ in terms of $\{|\chi\rangle,|\psi\rangle,[\chi|,[\psi| \}$, up to an irrelevant little group scale $\bar{\omega}$ that cancels from the final result.
Using,  $\beta=1+\frac{|q|}{2m_2^2}(|q|\mp\sqrt{|q|^2+4m_2^2})$, as well as the solutions \eqref{eq:onshell_sol} for the on-shell conditions for the external momenta,  to leading order in $|q|$, we find \footnote{Here one is free to choose either of the solutions for the quadratic equations. This in turn provides a  remarkable simplification in the computation as one does not have to average  between the positive and negative solutions, as it is customarily done when computing  the amplitude coefficients via  unitarity methods  \cite{Forde:2007mi,Badger:2008cm,Bern:2020buy,Chen:2021kxt,Aoude:2022trd}   }
\be\label{eq:spinormap}
\begin{split}
   |\lambda\rangle&=\bar{\omega} \sqrt{|q|}(|\chi\rangle+|\psi\rangle) +\mathcal{O}(q^{3/2})\,,\,\,\,\,
   [\lambda|=\frac{\sqrt{2}|q|}{\bar{\omega}}([\chi|-[\psi|) +\mathcal{O}(q^{3/2})\,,\\
   |\eta\rangle& = -\frac{m_2 \bar{\omega}}{\sqrt{|q|}}|\chi\rangle+ \mathcal{O}(q^{1/2})\,,\,\,\,\,
   [\eta| = \frac{m_2 }{\sqrt{|q|\bar{\omega}}}[\chi|+ \mathcal{O}(q^{1/2})\,.
\end{split}
\ee
These solutions  can be replaced into \eqref{eq:kinematics HCL gravitons} to obtain analog expressions for the   internal graviton variables. Explicitly, in the tetrad basis, to leading order in $|q|$, the internal graviton momenta take the form
\begin{equation}\label{eq:massless}
\begin{split}
    k_2^\mu& = \frac{|q|}{2\sigma v}\big[\sqrt{z^2-1} (u^\mu-\sigma \frak{t}^\mu )+\sigma v(\frak{x}^\mu+iz \frak{y}^\mu)    \big]\,,\\
  k_3^\mu& =- \frac{|q|}{2\sigma v}\big[\sqrt{z^2-1}(u^\mu-\sigma \frak{t}^\mu )-\sigma v(\frak{x}^\mu-iz \frak{y}^\mu)    \big]\,, \\
  w^\mu &=-\frac{|q|}{2\sigma (1-v z)}\Big[(u^\mu-\sigma \frak{t}^\mu )z+\sigma v(\sqrt{z^2-1}\frak{t}^\mu  +i (z^2-1)\frak{y}^\mu) \Big]\,,
\end{split}
\end{equation}
where the incoming 4-velocity  $u^\mu=p_1^\mu/m_1$. Here we have used \eqref{eq:wdef} as a definition for $w^\mu$, and the gauge \eqref{eq:wdef}, together with the Lorentz products 
 $p_1{\cdot}k_2=-\frac{1}{2}m_1 |q|\sigma v \sqrt{z^2-1} $ and $p_1{\cdot}\epsilon_2=-\frac{1}{2[23]}m_1|q|\sigma (1+v z) $ entering in \eqref{eq:wdef}. We can further identify the momentum transfer $q^\mu=|q|\frak{x}^\mu$, whereas $\frak{y}^\alpha=\varepsilon^{\delta\alpha\beta\gamma}\frak{t}_\delta \frak{x}_\beta \frak{z}_\gamma $, since $\varepsilon_{\alpha\beta\gamma\delta}\frak{t}^\alpha\frak{x}^\beta \frak{y}^\gamma\frak{z}^\delta =1$, with $\varepsilon^{\alpha\beta\gamma\delta}$ the four-dimensional Levi-Civita symbol. In terms of the external momenta \eqref{eq:external_tetrad}, and with $\frak{y}^\alpha = \frac{1}{m_1 m_2 \sigma v |q|} \mathcal{E}^\alpha$, where $ \mathcal{E}^\alpha$ was defined bellow \eqref{eq:spin_op}, 
 the massless momenta \eqref{eq:massless} recover the main text   expressions \eqref{eq:spin_op}. 

\section{Appendix B: Compton Amplitude from Teukolsky  solutions }

The explicit form of the amplitude \eqref{eq:ansatzspin}, contains the function of contact deformations $P_\xi$ \cite{Bautista:2022wjf}:

\begin{align}\label{eq:ansatz}
    P_\xi =& \sum_{m=0}^2 \xi^{m-1} (w\cdot a)^{4-2m}(w\cdot a- k_2\cdot a)^m (w\cdot a + k_3 \cdot a)^m r^{(m)}_{|a|}(k_2\cdot a, -k_3 \cdot a, w\cdot a) \nonumber \\ 
   & +\sum_{m=0}^{\infty} \left[ \frac{(w\cdot a)^{2m+6}}{\xi^{m+2}} p^{(m)}_{|a|}(k_2\cdot a, -k_3 \cdot a, w\cdot a) \right. \nonumber \\
    &\left. + \xi^{m+2} (w\cdot a- k_2\cdot a)^{m+3}(w\cdot a + k_3 \cdot a)^{m+3}\,q^{(m)}_{|a|}(k_2\cdot a, -k_3 \cdot a, w\cdot a) \right]\,,
\end{align}

where the multivariable polynomials $p^{(m)}_{|a|},q^{(m)}_{|a|},r^{(m)}_{|a|}$  up to order $a^6$  have the explicit form:
\begin{equation}
   \begin{split}r_{|a|}^{(m)} & =c_{1}^{(m)}+c_{2}^{(m)}(k_{2}{\cdot}a-k_{3}{\cdot}a)+c_{3}^{(m)}w{\cdot}a+c_{4}^{(m)}|a|\omega\\
 & \quad+c_{5}^{(m)}(w{\cdot}a-k_{2}{\cdot}a)(w{\cdot}a+k_{3}{\cdot}a)\\
 & \quad+c_{6}^{(m)}(2w{\cdot}a-k_{2}{\cdot}a+k_{3}{\cdot}a)w{\cdot}a\\
 & \quad+c_{7}^{(m)}(2w{\cdot}a-k_{2}{\cdot}a+k_{3}{\cdot}a)^{2}+c_{8}^{(m)}(w{\cdot}a)^{2}\\
 & \quad+c_{9}^{(m)}(k_{2}{\cdot}a-k_{3}{\cdot}a)|a|\omega+c_{10}^{(m)}w{\cdot}a|a|\omega+\mathcal{O}(a^{3})
\end{split}
\label{eq:polr}
\end{equation}
\begin{equation}
    p_{|a|}^{(m)}=d_{1}^{(m)}+\mathcal{O}(a)\,,\qquad q_{|a|}^{(m)}=f_{1}^{(m)}+\mathcal{O}(a)\,.
    \label{eq:polpq}
\end{equation}
Here $c_i^{m},d_i^{(m)}$ and $f_i^{(m)}$ are free coefficients that can be fixed either by imposing spin-shift symmetry arguments, or by matching the explicit BHPT computation. The different choices with their respective explicit values are indicated in \cref{tab:Teukolskysolutions}.

\begin{table*}[ht]
\resizebox{\textwidth}{!}{
\begin{tabular}{|c|c|c|c|c|c|c|}

\cline{1-4} \cline{2-4} \cline{3-4} \cline{4-4} \cline{6-7} \cline{7-7} 
Spin & Spurious-pole & Free Coeffs. & Teukolsky Solutions &  & Spin-Shift-Sym. & Free Coeffs.\tabularnewline
\cline{1-4} \cline{2-4} \cline{3-4} \cline{4-4} \cline{6-7} \cline{7-7} 
\hline
\hline
$a^{4}$ &  & $c_{1}^{(i)},\,\,i=0,1,2$ & $c_{1}^{(i)}=0,\,\,i=0,1,2$ &  & $\begin{aligned}c_{1}^{(i)} & =0,\,i=1,2\end{aligned}
$ & $c_{1}^{(0)}$\tabularnewline
\cline{1-4} \cline{2-4} \cline{3-4} \cline{4-4} \cline{6-7} \cline{7-7} 
\hline
\hline
$a^{5}$ & $c_{3}^{(2)}=4/15-c_{3}^{(0)}+c_{3}^{(1)}$ & $\begin{aligned}c_{2}^{(i)} & ,\,\,i=0,1,2\\
c_{3}^{(i)} & ,\,\,i=0,1\\
c_{4}^{(i)} & ,\,\,i=0,1,2
\end{aligned}
$ & $\begin{aligned}c_{2}^{(i)} & =0,\,\,i=0,1,2\\
c_{3}^{(0)} & =\alpha\frac{64}{15}\,,\,c_{3}^{(1)}=\alpha\frac{16}{3}\,,\\
c_{3}^{(2)} & =\frac{4}{15}(1+4\alpha)\,,\\
c_{4}^{(0)} & =\eta\alpha\frac{64}{15}\,,\\
c_{4}^{(1)} & =\eta\alpha\frac{16}{5}\,,\,c_{4}^{(2)}=\eta\frac{4}{15}
\end{aligned}
$ &  & $\begin{aligned}c_{j}^{(i)} & =0,\,\,i=1,2,\,\,j=2,3\\
c_{3}^{(0)} & =\frac{4}{15},\,\,c_{4}^{(i)}=0\,,i=0,1,2
\end{aligned}
$ & $c_{2}^{(0)}$\tabularnewline
\cline{1-4} \cline{2-4} \cline{3-4} \cline{4-4} \cline{6-7} \cline{7-7} 
\hline
\hline
$a^{6}$ & $\begin{aligned}c_{10}^{(2)} & =c_{10}^{(1)}-c_{10}^{(0)}\\
d_{1}^{(0)}=&-\frac{8}{45}+4 \sum_{i=0}^{2}(-1)^i c_7^{(i)}\\
&+\sum_{j=5}^6\sum_{i=0}^{2}(-1)^i c_j^{(i)}\\
f_{1}^{(0)}  =&\frac{4}{45}\\
& +\sum_{i=0}^2\sum_{j\in\{6,8\}}(-1)^i c_j^{(i)}
\end{aligned}
$ & $\begin{aligned}c_{5}^{(i)} & ,\,\,i=0,1,2\\
c_{6}^{(i)} & ,\,\,i=0,1,3\\
c_{7}^{(i)} & ,\,\,i=0,1,2\\
c_{8}^{(i)} & ,\,\,i=0,1,3\\
c_{9}^{(i)} & ,\,\,i=0,1,2\\
c_{10}^{i} & ,\,\,i=0,1
\end{aligned}
$ & $\begin{aligned}c_{j}^{(i)} & =0,\,\,i=0,1,2,\,\,j=5,7\\
c_{6}^{(0)} & =\alpha\frac{128}{45}\,,\,c_{6}^{(1)}=\alpha\frac{32}{9}\,,\\
c_{6}^{(2)} & =\frac{8}{45}(1{+}4\alpha)\,,c_{8}^{(0)}={-}\alpha\frac{512}{45},\,\\
c_{8}^{(1)} & ={-}\alpha\frac{160}{9},\,c_{8}^{(2)}={-}\frac{16}{45}(1{+}19\alpha),\\
c_{9}^{(0)} & =-\eta\alpha\frac{128}{45}\,,\,c_{9}^{(1)}=-\eta\alpha\frac{32}{15}\,,\\
c_{9}^{(2)} & =-\eta\frac{8}{45},\\
c_{10}^{(0)} & =-\eta\alpha\frac{256}{45},\\
c_{10}^{(1)} & =-\eta\alpha\frac{352}{45},c_{10}^{(2)}=-\eta\alpha\frac{32}{15}\\
d_{1}^{(0)} & =0\,,\,f_{1}^{(0)}=-\frac{4}{45}(1+4\alpha)
\end{aligned}
$ &  & $\begin{aligned}c_{j}^{(i)} & =0,\,\,i=0,1,2,\,\,j=5,9,10\\
c_{j}^{(i)} & =0,\,\,i=1,2,\,\,j=6,7,8\\
c_{8}^{(0)} & =-\frac{4}{45}-c_{6}^{(0)}\\
f_{1}^{(0)} & =0\\
d_{1}^{(0)} & =-\frac{8}{45}+c_{6}^{(0)}+4c_{7}^{(0)}
\end{aligned}
$ & $c_{6}^{(0)},c_{7}^{(0)}$\tabularnewline
\cline{1-4} \cline{2-4} \cline{3-4} \cline{4-4} \cline{6-7} \cline{7-7} 
\end{tabular}}
\caption{Second column:. Spurious pole cancellation constraints on the Compton ansatz.  Third column: Free contact coefficients after imposing the constraint of the second column. Fourth column: Match to  low-energy    Teukolsky solutions. Here $\alpha=1$, is a coefficient tracking  non-rational (digamma functions) contributions, and $\eta = 0,\pm1$, is a parameter that keeps track of the \textit{regular} ($\eta=0$), and \textit{exotic} ($\eta=\pm1$) pieces of the  scattering amplitude. Fifth column: Spin-shift symmetry constraints on the Compton ansatz. Sixth column: Free parameters for a spin-shift symmetric Compton ansatz.  Table recreated from \cite{Bautista:2022wjf}. }
\label{tab:Teukolskysolutions}

\end{table*}

\end{document}